\documentclass[aps,pra,twocolumn,showpacs,amsmath,amssymb,superscriptaddress,longbibliography, superscriptaddress]{revtex4-1}

\usepackage{graphicx}
\usepackage{verbatim}
\usepackage{epstopdf}
\usepackage{bm}
\usepackage{times}
\usepackage{comment}
\usepackage{dsfont}
\usepackage{color}
\definecolor{blue}{RGB}{0,0,255}
\usepackage[colorlinks,citecolor=blue,linkcolor=blue,urlcolor=blue]{hyperref}
\pdfoutput=1

\def\BE{\begin{equation}}
\def\EE{\end{equation}}
\def\BY{\begin{eqnarray}}
\def\EY{\end{eqnarray}}
\def\BI{\begin{itemize}}
\def\EI{\end{itemize}}
\def\L{\label}

\def\({\left (}
\def\){\right)}
\def\[{\left [}
\def\]{\right]}
\def\<{\langle}
\def\>{\rangle}

\def\BA{\begin{array}}
\def\EA{\end{array}}

\def\dd{\delta}
\def\D{\Delta}
\def\w{\omega}
\def\W{\Omega}

\def\d{\partial}
\def\g{\gamma}

\def\t{\tau}

\def\+{\dag}
\def\8{\infty}

\def\l{\lambda}
\def\ii{\text{i}}
\def\={\approx}

\def\->{\rightarrow}

\newcommand{\ud}{\,\mathrm{d}} 

\def\inn{|\textrm{in}\rangle}

\def\inn{\text{in}}

\def\um{$\rm{\mu m}$}
\def\um1{$\rm{\mu m}^{-1}$}

\def\Vn{V_n}

\def\Wn{W_n}

\usepackage{svg}

\usepackage{cancel}
\usepackage{enumitem}
\graphicspath{{images/}}

\allowdisplaybreaks[1] 

\usepackage{indentfirst}

\usepackage[normalem]{ulem}  

\usepackage[breakable]{tcolorbox}

\begin{document}

\title{
Effect of group-velocity dispersion on the generation of multimode pulsed squeezed light \\
in a synchronously pumped optical parametric oscillator
}

\author{Valentin Averchenko}
\email{valentin.averchenko@gmail.com}
\affiliation{St. Petersburg State University, 7/9 Universitetskaya Nab., 199034 St. Petersburg, Russia}

\author{Danil Malyshev}
\affiliation{St. Petersburg State University, 7/9 Universitetskaya Nab., 199034 St. Petersburg, Russia}

\author{Kirill Tikhonov}
\affiliation{St. Petersburg State University, 7/9 Universitetskaya Nab., 199034 St. Petersburg, Russia}

\date{\today}

\begin{abstract}

Parametric down-conversion in a nonlinear crystal is a widely employed technique for generating quadrature squeezed light with multiple modes, which finds applications in quantum metrology, quantum information and communication. 
Here we study the generation of temporally multimode pulsed squeezed light in a synchronously pumped optical parametric oscillator (SPOPO) operating below the oscillation threshold, while considering the presence of non-compensated intracavity group-velocity dispersion.
Based on the developed time-domain model of the system, we show that the dispersion results in mode-dependent detuning of the broadband supermodes of the pulsed parametric process from the cavity resonance, as well as linear coupling between these supermodes. 
With the perturbation theory up to the second order in the coupling coefficients between modes, we obtained a solution for the supermode amplitudes given an arbitrary number of modes and pump level.
The dispersion affects the quantum state of the supermodes by influencing their squeezing level and the rotation of the squeezing ellipse.
It also affects the entanglement among the supermodes, leading to reduced suppression of shot noise level as measured in the balanced homodyne detection scheme.
Furthermore, our study highlights the potential of  SPOPO with  group-velocity dispersion as a testbench for experimental investigations of multimode effects in linearly evanescent coupled parametric oscillators.

\end{abstract}

\maketitle

\section{Introduction}

Multimode squeezed light \cite{doi:https://doi.org/10.1002/9781119009719.ch5} is an essential resource for quantum networks \cite{crisafulli_squeezed_2013,kaiser_fully_2016}, boson sampling \cite{zhong_phase-programmable_2021,madsen_quantum_2022}, and optical quantum computation with continuous variables \cite{asavanant_generation_2019,walmsley_light_2023}. 
Additionally, it offers significant advancements in quantum metrology \cite{lawrie_quantum_2019, kamble_quantum_2024, conlon_verifying_2024}, timing measurement \cite{gosalia_leo_2023}, quantum magnetometry \cite{troullinou_squeezed-light_2021,wu_quantum_2023} and particularly, in gravitational wave detection \cite{aasi_enhanced_2013, oelker_squeezed_2014,page_gravitational_2021}.

Parametric down-conversion is a highly versatile and widely-utilized method for generating squeezed light in a variety of configurations and regimes. These include continuous or pulsed regimes, as well as different setups such as optical parametric oscillators with cavity configurations, nonlinear waveguides, integrated optics, multimode fibers, and more  \cite{wu_generation_1986, bachor_guide_2019}. 
The use of different configurations allows for control over the spatio-temporal characteristics and mode structure of the generated squeezed light \cite{fabre_modes_2020, piccardo_roadmap_2022, sukharnikov_managing_2021, kouadou_spectrally_2023, manukhova_noiseless_2017}. This provides flexibility in tailoring the properties of squeezed light to suit specific applications and experimental requirements.

It has been shown theoretically and experimentally that a synchronously pumped optical parametric oscillator (SPOPO) generates a quantum squeezed light with multiple time-frequency modes \cite{Patera2010,Roslund2014}.
In the time domain, the generated light exhibits a regular train of femtosecond correlated pulses. Each pulse is composed of a set of broadband supermodes, with each supermode being in a quantum independent squeezed state.
In the frequency domain, the generated light manifests as a set of frequency combs, with each comb also being in a quantum independent squeezed state.
This type of light possesses unique properties that make it highly attractive for a wide range of applications.
One of the applications that takes advantage of the temporal characteristics of short optical pulses with reduced quantum noise is the quantum improvement of time transfer between remote clocks \cite{lamine_quantum_2008,wang_sub-shot-noise_2018}. Another application is the quantum-enhanced measurement of the central frequency, mean energy, and spectral bandwidth of a light field \cite{cai_quantum_2021}.
Furthermore, the presence of thousands of quantum correlated frequency modes in the generated light can serve as a resource for generating cluster states, which are essential for all-optical multiplexed measurement-based quantum computing \cite{menicucci_fault-tolerant_2014,walschaers_photon-subtracted_2019} and quantum networks \cite{walschaers_emergent_2023, renault_experimental_2023, roman-rodriguez_multimode_2023}.

Understanding the details of light generation in SPOPO is crucial for controlling the characteristics of the generated light, improving the level of squeezing, and controlling the squeezed modes.
In SPOPO, the femtosecond supermodes generated experience group velocity dispersion inside the cavity caused by  the nonlinear crystal and cavity elements dispersion.
The dispersion length is a key parameter that characterizes the distance over which a pulse broadens due to material dispersion \cite{AkhmanovBook}. It is defined as the distance at which the pulse temporal width increases by a factor of 
$\sqrt{2}$. For a pulse durations on the order of hundreds of femtoseconds, the dispersion length is typically on the order of tenths of a millimeter for nonlinear materials and glasses.
Efforts are made to compensate for the dispersion by introducing additional passive cavity elements to the system.
However, in this study, we investigate the effect of non-compensated dispersion on the multimode quantum properties of the squeezed light generated in SPOPO below the oscillation threshold.

This paper is organized as follows. 
In Sec.~\ref{Sec2}, we present the time-domain model of SPOPO with intracavity group-velocity dispersion and the Heisenberg-Langevin equation that describes the quantum dynamics of a signal pulse circulating inside the oscillator's cavity.
In Sec.~\ref{Sec3}, we consider the system in the basis of supermodes of the pulsed parametric process. We show that it is equivalent to multiple optical parametric oscillators, and group-velocity dispersion causes detuning of supermode oscillators from their resonances and linear evanescent coupling between them.
In Sec.~\ref{Sec4}, we find an analytical solution for the amplitudes of oscillators using perturbation theory with respect to the coupling coefficients between them up to the second order.
In Sec.~\ref{Sec5}, we model time-domain pulsed balanced homodyne detection of the multimode light generated in SPOPO with group-velocity dispersion and derive an analytical expression for the spectrum of the detected difference photocurrent.
In Sec.~\ref{Sec6}, we numerically estimate the effects of intracavity group-velocity dispersion on the spectrum of the shot-noise suppression for a particular case of Hermite-Gaussian supermodes.
Sec.~\ref{Sec7} concludes the paper.

\section{Time-domain model of SPOPO with intracavity group-velocity dispersion}
\L{Sec2}
    \begin{figure}[htbp]
    \centering
    \includegraphics[width=1.00\columnwidth]{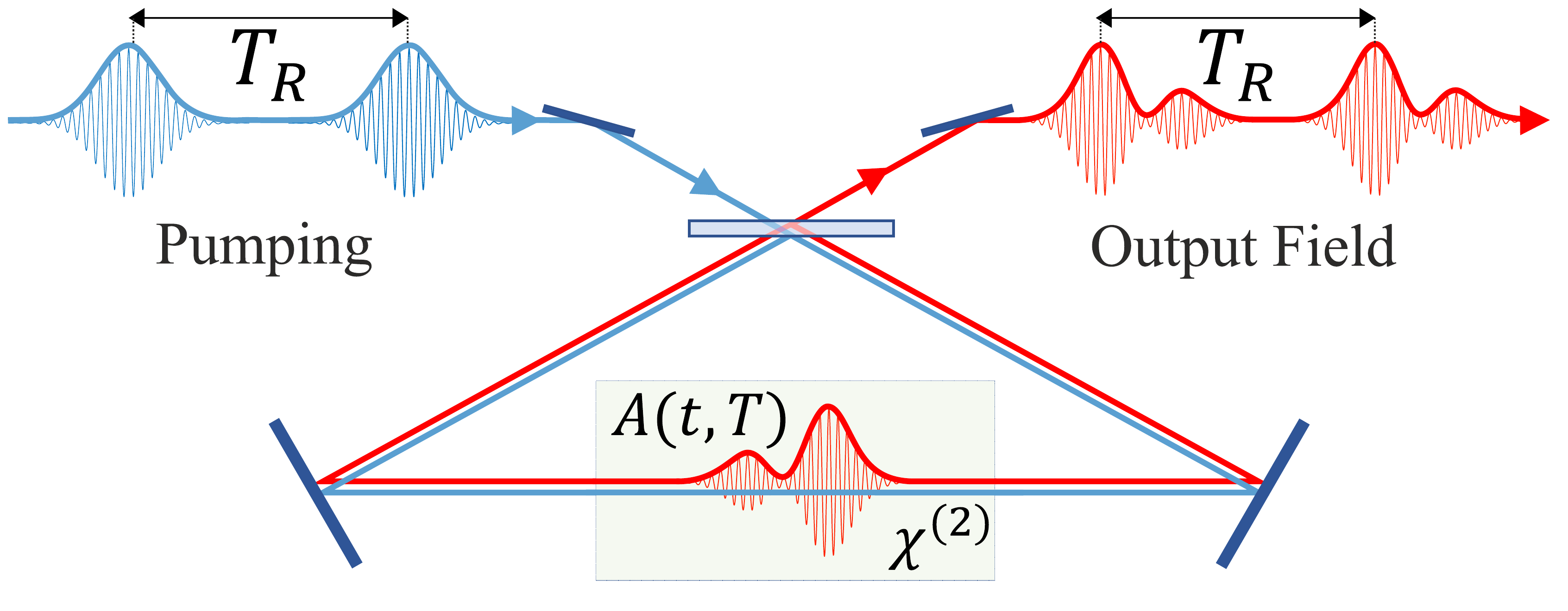}
    \caption{Principal scheme of the Synchronously Pumped Optical Parametric Oscillator (SPOPO)}
    \label{SPOPO_scheme}
    \end{figure}

The generation of a signal light in the SPOPO, shown in Fig.~\ref{SPOPO_scheme}, is modelled as follows in the time domain \cite{averchenko_quantum_2011, jiang_timefrequency_2012}.
A pump pulse enters the cavity through the coupling mirror and undergoes parametric down conversion process inside a nonlinear crystal.
As the generated signal pulse propagates within the cavity, a portion of the light exits the cavity at the coupling mirror, while the remaining portion reaches the crystal in synchronization with the arrival of the subsequent pump pulses.
During its propagation within the cavity, the signal pulse is subject to dispersive pulse broadening and chirping, which arise from group-velocity dispersion of the nonlinear crystal and the optical elements within the cavity. 
Also a vacuum noise is added to the signal on the coupling mirror. 
We consider the case when pump is not stored inside the cavity.
The model can be described using a Heisenberg-Langevin equation for the quantum mechanical operator (we omit notation of operators with the hat for brevity) of the slowly varying amplitude of the signal pulse $A(t, T)$ circulating inside the cavity. The equation reads under a number of approximations (see Appendix \ref{AppendixB} for the derivation):
	\begin{multline}
	\frac{\partial A(t,T)}{\partial T} 
	= \(-\frac{\g}{2} + \ii \Delta - \ii D \frac{\partial^2}{\partial t^2}\) A(t,T)\\
    + \frac{1}{2}\int G(t,t') A^\+(t',T) \ud t' +  F(t,T).
        \L{eqHL}
	\end{multline}
The equation is similar to the master equation used in the modeling of mode-locked lasers \cite{haus_mode-locking_2000, rana_quantum_2004, perego_coherent_2020}.
The operator $A (t,T)$ obeys the standard bosonic commutation relations, i.e. $\[A(t,T), A^\+(t',T') \] = T_R \delta(t-t') \delta(T-T')$. 
The additional time variable $T$ describes the evolution of the pulse over time scales larger than the cavity round-trip time $T_R$, i.e. $A(t, T=NT_R)$ gives amplitude of the pulse after $N$ round-trips.  
The operator is normalized such that $\< A^\+(t,T) A(t,T)\>$ gives the photon number flux.
The cavity field decays with the rate $\gamma$  due to a leakage at the coupling mirror. After $(\g T_R)^{-1} = {\cal F}/2\pi$ round trips in the cavity pulse intensity decays $e$-times, where $\cal F$ is the cavity finesse. Also, $\g/2\pi$ defines full width at half maximum of the Lorentzian spectral line shape of the cavity resonance.
The model takes into account a detuning $\Delta$ of the central frequency of the signal field, which is a half of the pump carrying frequency in the degenerate parametric process, from the resonance. $\D \times T_R$ gives a phase shift of the pulse upon round-trip in the cavity.
Dynamics of SPDC process in the nonlinear crystal is described with an integral kernel $G(t,t')$, which depends on the crystal characteristics (geometry, nonlinear propoerties, dispersion) and temporal and spatial characteristics of the pump pulse (see Appendix~\ref{AppendixA}). The kernel is complex valued and symmetric in general.
Group-velocity dispersion is quantified via a parameter  $D$.
$F(t,T)$ is a Langevin noise with the only non-zero correlation function $\<F(t,T) F^{\+}(t',T')\> = \g\dd(t-t') \dd(T-T')$.
The following expression relates the signal field inside the cavity with the output field: 
	\begin{align}
	B(t,T) = \sqrt{\gamma T_R} \( A(t,T) - \g^{-1} F(t,T)\).
	\end{align}

\section{Effect of Intracavity Dispersion on the Supermodes of the Parametric Process}
\L{Sec3}

    \begin{figure}[htbp]
    \centering
    \includegraphics[width=1.0\columnwidth]{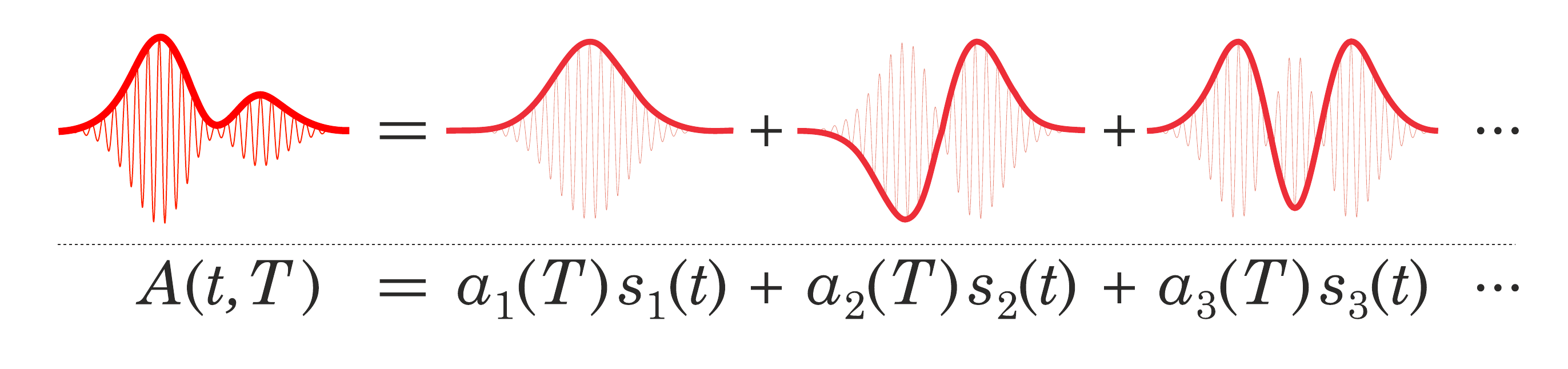}
    \caption{Amplitude of intracavity signal pulse $A(t,T)$ can be decomposed as a sum of amplitudes of supermodes $\{a_n(T)\}$ with temporal profiles $\{s_n(t)\}$, which  are the Hermite-Gaussian functions in this particular case.}
    \label{fig_decomposition}
    \end{figure}
\begin{figure}[htbp]
    \centering
    \includegraphics[width=\columnwidth]{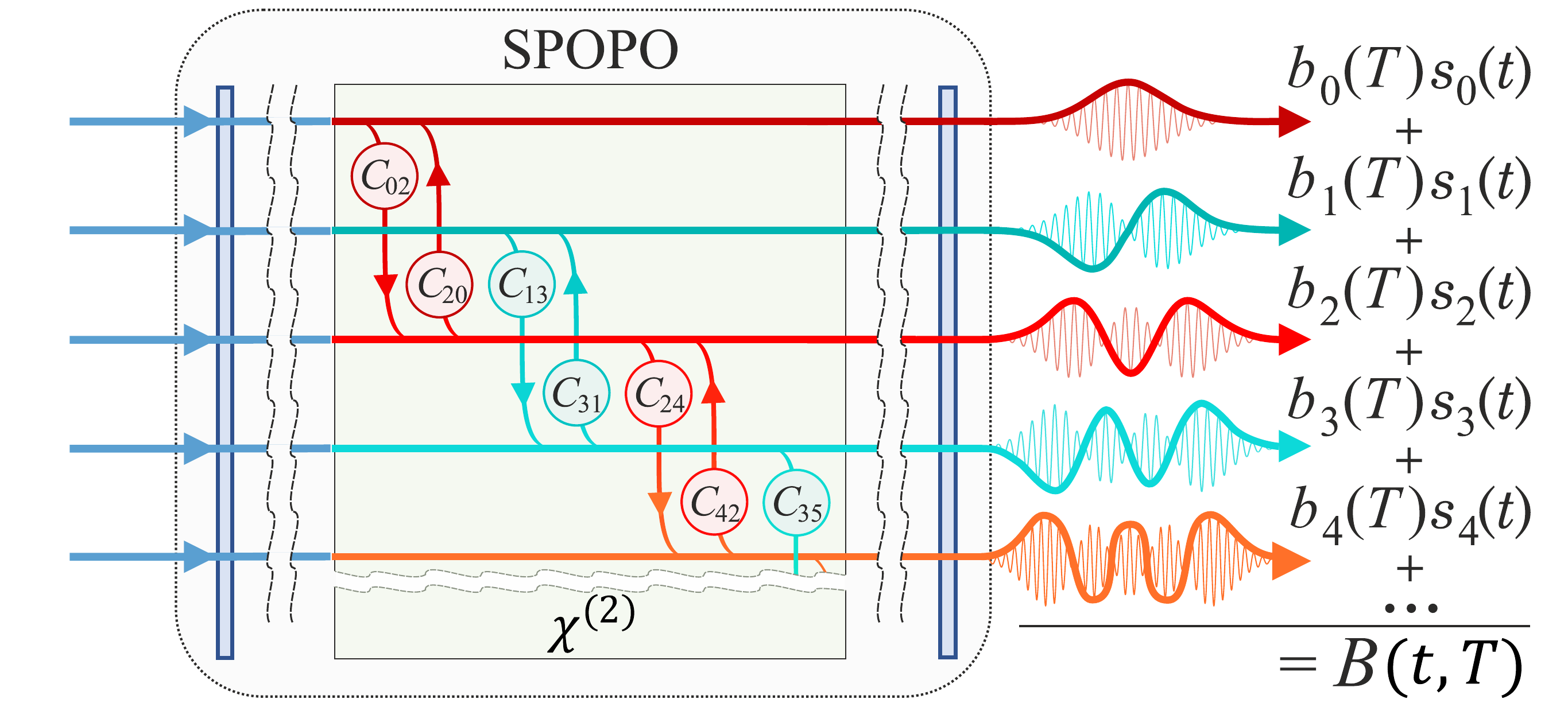}
    \caption{
    SPOPO can be represented as a set of parametric oscillators in the basis of broadband supermodes of the pulsed parametric process. 
    Each oscillator corresponds to a pulse with the supermode temporal envelope circulating inside the cavity. 
   Group-velocity dispersion causes two effects: the shift of the resonance frequency of each supermode oscillator $\D_n$; the linear coupling between oscillators with coefficients $C_{nm}$. On the figure a particular case of coupling between the neighbour modes of the same parity is shown, which holds, specifically for the Hermite-Gaussian broadband supermodes (\ref{HG_def}).
    } 
    \label{fig_coupling}
    \end{figure}

The integral kernel of SPDC $G(t,t')$ which is complex and symmetric in general admits the decomposition (for example, see \cite{arzani_versatile_2018, horoshko_bloch-messiah_2019}):
	\begin{align}
	G(t,t') = \sum_{n=0,1,2,\ldots} \l_n s_n(t) s_n(t'). 
	\end{align}
Here the decomposition coefficients $\{\l_n\}$ are real and non-negative; the set of orthonormal functions $\{s_n(t)\}$ which are complex in general is full. The functions are defined by the equations:
	\begin{align}
	\int G(t,t') s_n^*(t') dt' = \l_n s_n(t). 
	\end{align}
The amplitude of the signal field inside the cavity can be decomposed using the set of functions $\{s_n(t)\}$ \cite{blow_continuum_1990, brecht_photon_2015, raymer_temporal_2020}:
	\begin{align}
    A(t,T) = \sum_{n=0,1,2,\ldots} a_n(T) s_n(t), 
    \L{super_decomp}
	\end{align}
where the decomposition coefficients $\{a_n(T)\}$ are the quantum mechanical operators:
	\begin{align}
    a_n(T) = \int A(t,T) s_n^*(t) \ud t.
    \L{operators_a}
	\end{align}
The decomposition (\ref{super_decomp}) shows that the signal pulse inside the cavity can be seen as a sum of pulses, also called "supermodes" \cite{Patera2010}, with the temporal profiles $s_n(t)$ and the complex amplitudes $a_n(T)$, which evolve upon circulation of pulses inside the cavity -- see Fig.~\ref{fig_decomposition}. 
The operators (\ref{operators_a}) satisfy the commutation relation $\[a_n(T), a_m^\+(T') \] = T_R \delta_{nm} \delta(T-T')$ and are normalized so that $\<a_n^\+(T) a_n(T)\>$ gives the mean number of photons in the supermode $n$.
The amplitudes of the supermodes obey the following system of equations derived from (\ref{eqHL}), using the orthogonal expansions for the amplitude of signal field inside the cavity (\ref{super_decomp}), and similar expansions for the amplitude of the Langevin source and of the output signal field:
	\begin{multline}
	\frac{\d a_n(T)}{\d T}
    =  \(-\frac{\g_n}{2} + \ii\Delta_n\)a_n(T) \\
    + \frac{\l_n}{2} a_n^\+(T) - \ii\sum_{m\neq n} C_{nm} a_m(T) + f_n(T)
	\L{coupled_modes_equations_T}
	\end{multline}
The equations show that the SPOPO can be viewed as a set of parametric oscillators, where each oscillator is characterized by the complex bosonic amplitude $a_n$, parametric gain $\l_n$, detuning from parametric resonance $\Delta_n$, loss rate $\g_n$, and a noise source $f_n$ caused by losses - see Fig.~\ref{fig_coupling}. As such the system generates squeezed states in the several supermodes $s_n(t)$ \cite{Patera2010, onodera_nonlinear_2022}.

The equations also show that group velocity dispersion inside the optical cavity causes two effects for the supermodes.
First, it is the linear coupling between modes, characterized with the mode-dependent complex coupling coefficient $C_{nm}$:
    \begin{align}
    & C_{nm} = D O_{nm}, \\
    & O_{nm} = \int s_n^*(t) \frac{d^2}{dt^2} s_m(t) \ud t.
    \L{overlap_def}
    \end{align}
Integrating by parts (\ref{overlap_def}) one can show that a matrix of coupling coefficients is Hermitian.
Furthermore, since the coupling coefficient is proportional to the overlap of a mode with the second-order derivative of another mode, then the coupling is non-zero only between modes of the same parity. It is because the second-order dispersion causes broadening and chirping of a pulse and does not change its symmetry. Summing up,
    \begin{align}
    & C_{nm}^* = C_{mn},\\
    & C_{\text{odd even}} = 0.
    \end{align}
Second, the dispersion causes a mode-dependent detuning $\D_n$ of the supermodes from cavity resonance:
    \begin{align}
    \D_n = \D - C_{nn}.
    \L{D_n_def}
    \end{align}
The detuning is caused by the mode-dependent phase accumulated by the each broadband supermode during every round trip inside the cavity, which is a temporal analog of the Gouy phase \cite{dioum_temporal_2023, horoshko_interferometric_2023}.
Both the detuning of supermodes and the coupling between them affects their quantum states that will be analyzed below.

It is worth to mention that we also generalized the model and assumed in the above equations that losses are mode dependent, i.e. $\gamma \rightarrow \g_n$, for example, due to different bandwidth of the supermodes.
Correlation functions for noise terms of each mode read accordingly:
    \begin{align}
    \<f_n(T) f_m^{\+}(T')\> = \g_n \dd_{nm} \dd(T-T').  
    \L{correlator_f_T}
    \end{align}
The input-output relation for the $n$-th mode reads:
    \begin{align}
    b_n(T) = \sqrt{\gamma_n T_R} \( a_n(T)-\frac{1}{\g_n} f_n(T)\). 
    \L{in_out_n}
    \end{align}

\section{Perturbative solution of equations for coupled supermodes}
\L{Sec4}

We solve coupled equations (\ref{coupled_modes_equations_T}) in the  frequency domain. Namely, for each mode we define a corresponding complex spectral amplitude as follows (we use the same symbol for the spectral amplitudes):
	\begin{align}
	a_n(\W) = \frac{1}{\sqrt{2\pi}} \int a_n(T) e^{\ii \W T} \ud T.
    \L{Fourier}
	\end{align}
The spectral amplitudes obey the commutation relation $[a_n(\W), a_m^\+(\W')] = T_R \delta_{nm} \delta(\W-\W')$.
They also satisfy the following algebraic equations, which are obtained from the differential equations (\ref{coupled_modes_equations_T}) by applying the Fourier transform to them
	\begin{multline}
	-\ii\W a_n(\W)
    = \(-\frac{\g_n}{2} + \ii\Delta_n\) a_n(\W) \\
    + \frac{\l_n}{2} a_n^\+(-\W) + 
	f_n(\W) - \ii\sum_{m\neq n} C_{nm} a_m(\W).
	\L{coupled_modes_equations}
	\end{multline}
Correlation functions for noise terms are obtained from (\ref{correlator_f_T}) using the Fourier transformation and read:
\begin{align}
    \<f_n(\W) f_m^{\+}(\W')\> = \g_n \dd_{nm}  \dd(\W-\W').
    \L{f_correlator}
\end{align}

We solve the coupled equations using the perturbation theory with respect to the coupling coefficients between modes $C_{nm}$ up to the second order (see Appendix \ref{AppendixB} for the details).
The solution can we written in the following form for the $n$-th supermode:
    \begin{multline}
    a_n(\Omega) = U_n(\Omega) f_n (\Omega) + V_n(\Omega) f_n^{\dagger} (-\Omega) 
    \\+ h_n(\Omega) + \kappa_n(\Omega).
    \L{solution}
    \end{multline}
The first line of the solution represents the two-mode squeezing transformation of vacuum noise at frequencies $\pm \W$ (see \cite{kolobov_spatial_1999}). 
The quantum state of these spectral components in the zeroth order of the perturbation theory is a two-mode pure squeezed vacuum state.
Squeezing coefficients $U_n, V_n$ depend on intracavity dispersion via mode-dependent resonant frequency $\Delta_n$ that affects characteristics of the squeezed state, such as degree of squeezing, orientation of the squeezing ellipse.
Squeezing coefficients read:
	\begin{align}
	& U_n(\W) = \frac{1}{H_n(\W)},\\
	& V_n(\W) = \frac{1}{H_n(\W)} \frac{\l_n}{2} \frac{1}{\frac{\g_n}{2}+\ii \D_n - \ii \W},\\
	& H_n(\W) = \frac{\g_n}{2}-\ii \D_n - \ii \W - \(\frac{\l_n}{2}\)^2 \frac{1}{\frac{\g_n}{2}+\ii\D_n-\ii\W}.
	\L{sol_detuning}
	\end{align}
The second line of the solution (\ref{solution}) is obtained as the first and the second order of the perturbation theory and represents an additional noise in each supermode due to its linear coupling with other modes.
This noise results in a mixed squeezed state of each supermode.
Expressions for these noise terms read:
    \begin{align}
    h_n(\Omega) = \sum_{m \neq n} \left[U_{nm}(\Omega) f_m(\Omega) + V_{nm}(\Omega) f_m^\dagger(-\Omega)\right],\\
    \kappa_n(\Omega) = \sum_{\substack{m \neq n \\ p \neq m}} \left[U_{nmp}(\Omega) f_p(\Omega) + V_{nmp}(\Omega) f_p^\dagger(-\Omega)\right],
    \end{align}
where
    \begin{equation}
    \begin{aligned}
        {U_{nm}(\Omega)} &= i [-U_n(\Omega) C_{nm} U_m(\Omega) + V_n(\Omega) C_{nm}^* V_m^*(-\Omega)],
        \\
        {V_{nm}(\Omega)} &= i [-U_n(\Omega) C_{nm} V_m(\Omega) + V_n(\Omega) C_{nm}^* U_m^*(-\Omega)],
        \\
        {U_{nmp}(\Omega)} &= i [-U_n(\Omega) C_{nm} U_{mp}(\Omega) + V_n(\Omega) C_{nm}^* V_{mp}^*(-\Omega)],
        \\
        {V_{nmp}(\Omega)} &= i [-U_n(\Omega) C_{nm} V_{mp}(\Omega) + V_n(\Omega) C_{nm}^* U_{mp}^*(-\Omega)].
    \end{aligned}
    \label{uv}
    \end{equation}
An Output amplitude of the $n$-th mode is obtained using the input-output relation (\ref{in_out_n}) and reads
    \begin{multline}
    b_n(\Omega)/\sqrt{\gamma_n T_R} =  W_n(\Omega) f_n (\Omega) + V_n(\Omega) f_n^{\dagger} (-\Omega) 
    \\+ h_n(\Omega) + \kappa_n(\Omega) ,
    \L{b_solution}
    \end{multline}
where $ W_n(\Omega) = U_n(\Omega) - {\gamma_n^{-1}}$ and $|\Wn(\W)|^2 - |\Vn(\W)|^2 = \g_n^{-2}$.	

It is worth stressing that while $C_{nn}$ and $C_{nm}$ are of the same order for SPOPO, the obtained solution is rigorous with respect to both $C_{nn}$ and the mode-dependent detuning $\D_{n}$ from the cavity resonance, while perturbative with respect to $C_{nm}$.

\section{Pulsed balanced Homodyne detection of supermodes}
\begin{figure}[htbp]
    \centering
    \includegraphics[width=\columnwidth]{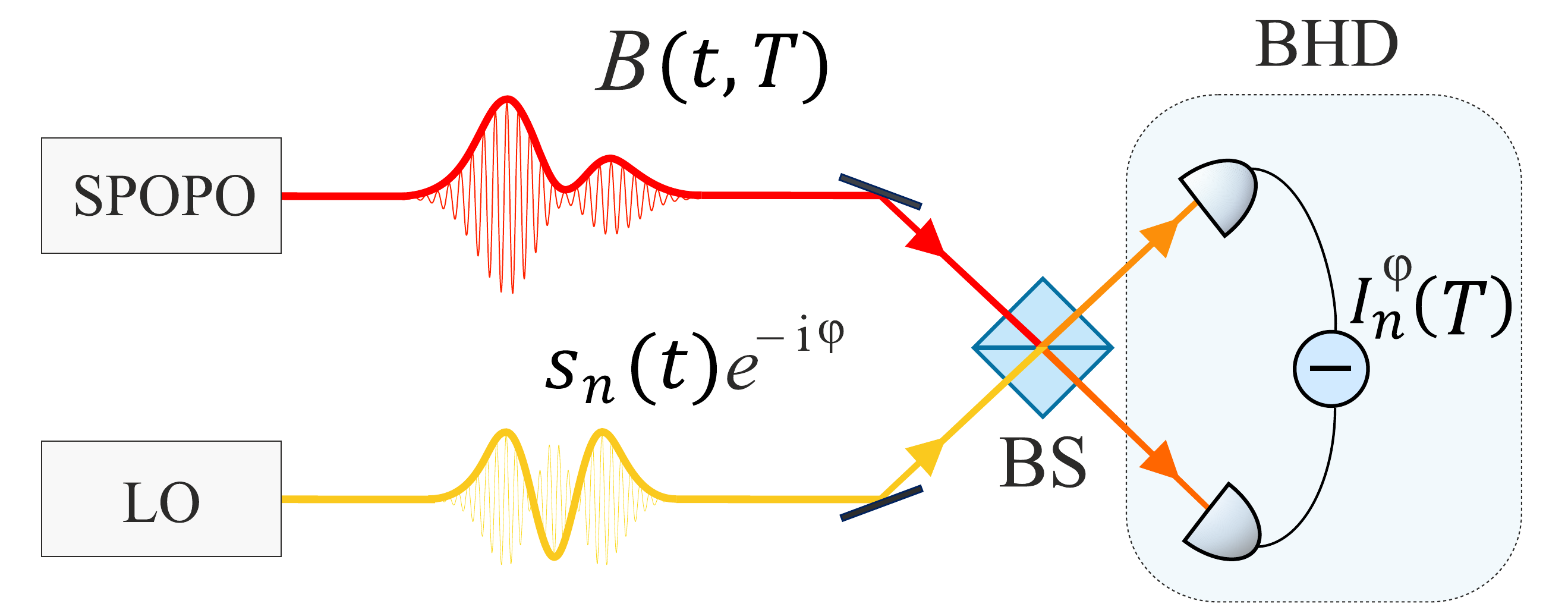}
    \caption{The principal scheme of balanced homodyne detection of pulsed light from SPOPO using a periodic pulsed local oscillator.
    The pulses of the local oscillator are shaped to match a desired supermode with the envelope $s_n(t)$. The relative phase shift $\varphi$ of the local oscillator can be controlled. 
    } 
    \label{BHD}
    \end{figure}

\L{Sec5}
The obtained above expressions allow to calculate an arbitrary characteristic of the generated signal field.
Here we consider balanced homodyne detection of the field and calculate the measured noise of the difference photocurrent.
The principal scheme of the measurement procedure is shown in Fig.~\ref{BHD}.
The signal field interferes on a symmetric beam splitter with a matched periodic pulsed field of the local oscillator, which has the same central frequency. The temporal shape of the local oscillator pulses can be controlled, as well as the global phase relative to the signal field. The intensities of the fields are measured at the outputs of the beam splitter using identical photodetectors, which we assume to be ideal with unit efficiency and no noise. Subtracting the two resulting photocurrents yields the difference photocurrent at the output of the measurement procedure.
We assume that the shape of the local oscillator pulses matches the shape of one of the temporal modes of the signal field $s_n(t)$ without intracavity chromatic dispersion.
We also assume temporal averaging over each single pulse.
This is a time-domain balanced detection \cite{smithey_measurement_1993, hansen_ultrasensitive_2001}.
The difference photocurrent at the output of the measurement system depends solely on the amplitude of the matched temporal mode:
	\begin{align}
	I_{n}^\varphi(T) \propto b_n(T) e^{\ii\varphi} + b_n^{\+}(T) e^{-\ii\varphi},
	\end{align}
where the phase $\varphi$ between signal and local oscillator fields can be controlled.
We assume further spectral analysis of the photocurrent. 
The spectral amplitude of the photocurrent, similar to (\ref{Fourier}), reads
	\begin{align}
	I_{n}^\varphi(\W) \propto b_n(\W) e^{\ii\varphi} + b_n^{\dag}(-\W) e^{-\ii\varphi}.
\label{Iw_BHD_def}
	\end{align}
Using results of the previous section we get the following expression for the photocurrent spectrum (see Appendix \ref{AppendixC}):
    \begin{multline}
    S_{n}^\varphi(\W) = \<I_n^\varphi(\W) I_n^{\varphi\+}(\W)\> \\
    \propto \gamma_n^2 \Big|W_n(\Omega) e^{\ii\varphi} + V_n^*(-\Omega)  e^{-\ii\varphi} \Big|^2 \\
    + \sum_{m \neq n} \gamma_n\gamma_m \Big|U_{nm}(\Omega)  e^{\ii\varphi} + V_{nm}^*(-\Omega)  e^{-\ii\varphi}\Big|^2\\
    + 2 \gamma_n^2 \mathbf{Re}\bigg\{ 
        \Big[W_n(\Omega)  e^{\ii\varphi} + V_n^*(-\Omega)  e^{-\ii\varphi}\Big] \\
    \times \sum_{m \neq n} \Big[U_{nmn}^*(\Omega)e^{-\ii\varphi} +V_{nmn}(-\Omega) e^{\ii\varphi}\Big]
    \bigg\}.
    \label{II2_total}
    \end{multline}
The expression after the proportionality symbol gives the photocurrent spectrum normalized to the shot noise level.
The first term in the expression represents the spectrum in the absence of coupling between supermodes, but in the presence of dispersion induced mode-selective detuning from cavity resonance.

\section{Photocurrent spectrum for balanced homodyne detection of Hermite-Gaussian supermodes}
\L{Sec6}

Here we analyze the photocurrent spectrum measured in the scheme of balanced homodyne detection for a specific case of the Hermite-Gaussian supermodes which are a good approximation of experimental results \cite{Patera2010, Roslund2014}. Normalized profiles of the modes in the temporal domain are given by:
	\begin{align}
	s_n(t) = i^n \( \sqrt{\pi} 2^n n! \t_s \)^{-\frac{1}{2}} H_n(t/\t_s) e^{- t^2/2 \t_s^2},
    \L{HG_def}
	\end{align}
where $H_n$ are Hermite polinomials. $\t_s$ defines temporal duration of supermodes and depends both on properties of pump pulses and the nonlinear crystal \cite{wasilewski_pulsed_2006, Patera2010}. For $n=0$ the supermode is the Gaussian function and the full width at half-maximum of its intensity profile, i.e. modulus squared, equals to $2\sqrt{\ln2} \t_s$.
The overlap integral for the supermodes (\ref{overlap_def}) reads 
\footnote{
It is worth mentioning that the matrix elements coincide with the matrix representation of the quantum mechanical operator for the squared momentum of a one-dimensional harmonic oscillator in the Fock states basis
$\<n| \hat p^2 |m\>$}:
\begin{multline}
    O_{nm} \t_s^2  = 
    -\left(
\begin{array}{ccccccc}
 \frac{1}{2} & 0 & \frac{1}{\sqrt{2}} & 0 & 0 & 0 & \cdots  \\
 0 & \frac{3}{2} & 0 & \sqrt{\frac{3}{2}} & 0 & 0 & \cdots \\
 \frac{1}{\sqrt{2}} & 0 & \frac{5}{2} & 0 & \sqrt{3} & 0 & \cdots \\
 0 & \sqrt{\frac{3}{2}} & 0 & \frac{7}{2} & 0 & \sqrt{5} & \cdots \\
 0 & 0 & \sqrt{3} & 0 & \frac{9}{2} & 0 & \cdots\\
 0 & 0 & 0 & \sqrt{5} & 0 & \frac{11}{2} & \cdots \\
 \vdots & \vdots & \vdots & \vdots & \vdots & \vdots & \ddots \\
\end{array}
\right) \\
= -\(n+\frac{1}{2}\) \dd_{n m} \\
 - \frac{\sqrt{(n-1)n}}{2} \dd_{n m+2} -  \frac{\sqrt{(n+1)(n+2)}}{2} \dd_{n m-2},
 \L{O_matrix}
\end{multline}
where $\dd_{nm}$ is the Kronecker delta.
Thus the coupling due to dispersion appears only between neighbor modes of the same symmetry schematically represented in Fig.~\ref{fig_coupling} and the coupling increases for higher order modes.
In turn, the coupling coefficients given by the expression (\ref{overlap_def}) are real valued.

\subsection{Analytical results}

We consider a limiting case of weak SPDC and small dispersion and perform series expansions of the photocurrent spectrum (\ref{II2_total}) with respect to $\l_n/\g_n, \D_n/\g_n \ll 1$ keeping first non-zero orders.
The photocurrent noise at $\W =0$ reads:
    \begin{multline}
    S_n^{\varphi = 0, \frac{\pi}{2}}(\W=0) \propto 1 \pm 4\frac{\l_n}{\g_n} \(1 - 12 \(\frac{\D_n}{\g_n}\)^2 \) \\
    \mp 16 \sum_{m = n \pm 2} \frac{C_{nm}^2}{\g_m\g_n}  \(2\frac{\l_n}{\g_n} + \frac{\l_m}{\g_m}\). 
    \L{analytical_SNL}
    \end{multline}
For supermodes with $\l_n \neq 0$ the expression shows maximal excess/suppression of the shot-noise at zeroth frequency for $\varphi = 0, \frac{\pi}{2}$, respectively.
The mode-dependent detuning from the cavity resonance $\D_n$ causes the reduction of the excess/suppression of the shot-noise by the factor $1 - 12 \D_n^2/\g_n^2$.
This is one of the effects of intracavity dispersion.
The last term in the expression (\ref{analytical_SNL}) describes the additional reduction of shot noise due to the coupling between modes.
This effect is proportional to both the gain of the mode $\l_n$ itself and to the gain of the modes to which it is coupled $\l_m$.

It also follows from the expression that even if a supermode has zero gain $\l_n = 0$, there will still be non-zero light power in the mode and non-zero difference photocurrent measured. This is because of the coupling of the mode to neighbouring modes with non-zero gain.
Otherwise such a supermode is not excited and remains in the vacuum state in the second order of the perturbation theory. 
The higher order terms will determine further propagation of excitation between the coupled modes, even if they have zero gain.
The result suggests that the coupling matrix (\ref{overlap_def}) can be truncated at the higher-order supermodes that are not excited by coupling or have significant losses, such that photons of the mode leave the cavity before leaking into the higher-order supermodes.

The above results also suggest that group-velocity dispersion can be used to resonantly amplify one of the $n$-indexed supermodes by tuning the cavity to the resonance of the mode such that $\Delta_n = \Delta - C_{nn} = 0$.
However, it also follows that achieving nondegeneracy of the modes comes at the cost of coupling between the neighboring modes, which leads to their entanglement and can reduce the purity of their quantum states.

    \begin{figure}[htbp]
    \centering
    \includegraphics[width=\linewidth]{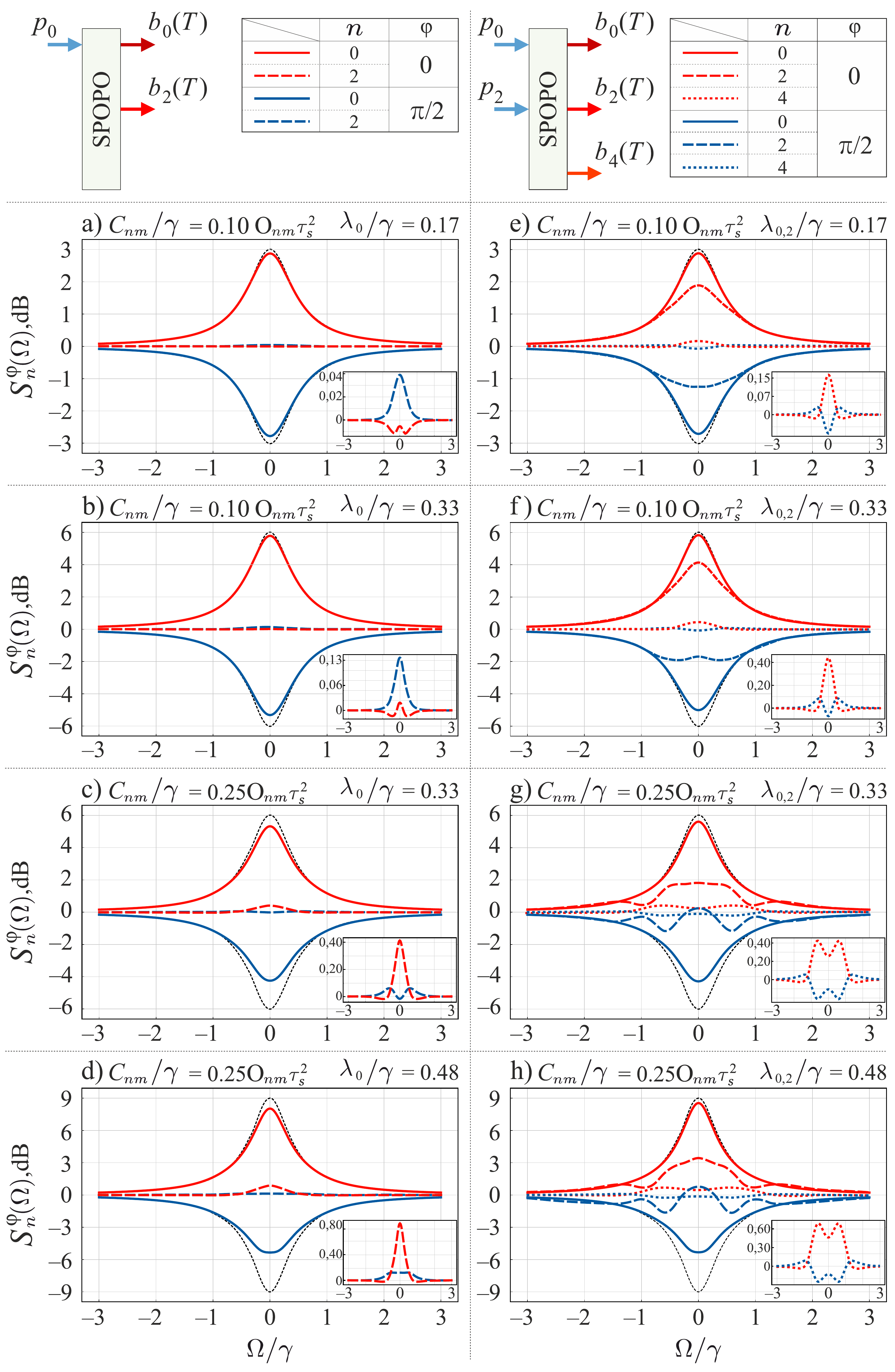}
    \caption{Modelling results for the photocurrent spectrum measured in the scheme of balanced homodyne detection of the pulsed signal light generated in SPOPO for different operation regimes of SPOPO and different values of the second-order dispersion. (Top) Excitation schemes of supermodes in SPOPO and color coding for plots. (Left) $0$th supermode is excited in SPOPO and $0$th and $2$nd supermodes are detected. (Right) $0$th and $2$nd supermodes are equally excited and $0$th, $2$nd, and $4$th supermodes are detected.
    Black dashed lines show the spectra of the photocurrent detected for the $0$th or $2$nd supermode in the absence of group-velocity dispersion in the cavity.
    The insets in the left and right columns show the photocurrent spectra when detecting the $2$nd or $4$th supermodes, respectively.
    } 
    \label{S_W}
    \end{figure}

\subsection{Numerical results}

We further analyze the impact of dispersion beyond the previous limit case (\ref{analytical_SNL}) by using the general expression for the photocurrent (\ref{II2_total}). This analysis incorporates experimental parameters from the study on the generation and measurement of multimode squeezed light \cite{Roslund2014}.
Spectral profiles of the generated broadband supermodes can be approximated by the Hermite-Gaussian functions with the full width at half maximum of the fundamental supermode equal to 8.5 nm that corresponds to the characteristic temporal duration $\t_s = 67$ fs (\ref{HG_def}).
Repetition rate of $T_R^{-1} = 76$ MHz and cavity finesse of ${\cal F} = 26$ result in the full width half maximum of the resonance of $\gamma/2\pi = 3$ MHz.
In turn, field intensity decays $e$-times after the following number of round trips $N_\gamma = (\g T_R)^{-1} = {\cal F}/2\pi = 8$.

Group-velocity dispersion causes temporal broadening of phase non-modulated optical pulses. 
For the Gaussian pulse with  duration $\t_s$ the effect can be quantified by the dispersion length $L_D = \t_s^2/k_2$, where $k_2 = {\partial^2 k(\w)}/{\partial \w^2}|_{\w=\w_s}$ represents group-velocity dispersion at the central frequency $\w_s$. The pulse broadens by a factor of $\sqrt{2}$ after propagating through a dispersive material with length $L_D$.
For a signal wavelength  $\l_s = 800$ nm BIBO crystal has $k_2 = 136$ fs$^2$/mm that results in $L_D = 33$ mm. 
For the dispersive material length $L = 2$ mm inside the cavity the pulse broadens by a factor of $\sqrt{2}$ after completing $N_D = L_D/L = 17$ round trips. 
Thus, one gets $N_\g/N_D \approx 0.5$, indicating that the generated signal pulse inside the cavity exits before the significant impact of dispersion. The effect of dispersion on the generated pulsed light can be considered as a perturbation under the given experimental conditions (see Appendix~\ref{AppendixD} for the validation of the perturbation approach).
Following the definition of $C_{nm}$ (\ref{overlap_def}) and $D$ (\ref{D_def}) one can show that for the Hermite-Gaussian supermodes
    \begin{align}
    \frac{C_{nm}}{\g} = \frac{\sqrt{\cal R}}{2} \frac{N_\g}{N_D} O_{nm} \t_s^2.
    \end{align}
Thus for the experimental parameters $C_{nm}/\g \approx 0.25 \times O_{nm} \t_s^2$.
In addition, amplitude reflection coefficient $\sqrt{\cal R}$ and loss rate $\gamma$ may depend on the supermode.

We consider regimes of SPOPO where one to five of the first supermodes are excited with different amplitudes.
This can be achieved by adjusting the spectral profile of the pump pulses and the parameters of the nonlinear crystal \cite{patera_quantum_2012, ansari_tailoring_2018, horoshko_few-mode_2024}.
In our analysis, we consider cases of 3, 6, 9 dB of squeezing for supermodes at zeroth frequency without the dispersion. 
By using expressions (\ref{solution}) and (\ref{sol_detuning}), one can show that such levels of squeezing are achieved for $\l_n/\g = 0.17, 0.33, 0.48$, respectively. We also assume equal loss rate for the supermodes.

Fig.~\ref{S_W}(a-d) shows the spectrum of the photocurrent, normalized to the shot noise level, when the zeroth supermode is excited.
The $0$th or the $2$nd mode is detected tailoring local oscillator pulses.
As expected from the analytical expression (\ref{analytical_SNL}),  group-velocity dispersion reduces the maximum quantum state squeezing or anti-squeezing of the excited mode, resulting in a smaller excess or suppression of the shot noise level upon mode detection.
There is also a non-zero power in the second mode due to its coupling to the zeroth mode, leading to a deviation of the photocurrent spectrum from the shot noise level.
Furthermore, the squeezing or anti-squeezing of the mode state is observed for the flipped phase of the local oscillator with respect to the $0$th mode.
These effects are stronger for sideband frequencies $\W/\gamma < 1$
 and become more pronounced with increased dispersion. They are also less influenced by the pump level.
Numerical results show that higher order modes are not excited, similar to the analytical result in the limiting case (\ref{analytical_SNL}). 
This is the second-order consequence of the perturbation theory: a mode that is pumped also excites its direct neighboring coupled modes; a mode with zero gain does not excite direct neighboring modes.
Fig.~\ref{S_W}(e-h) shows the spectrum of the photocurrent when the $0$th and  the $2$nd supermodes are equally excited simultaneously.
The $0$th, the $2$nd and the $4$th supermodes are detected tailoring local oscillator pulses.
Comparing with Fig.~\ref{S_W}(a-d) one concludes that for the chosen parameters the $0$th supermode is slightly affected by the excitation of the $2$nd supermode: there is a small reduction of squeezing and some increase of anti-squeezing for the chosen phases of the local oscillator. 
The $2$nd mode shows a significant deviation of the spectrum from a Lorentzian profile of the photocurrent because the dispersion effect is stronger for higher order supermodes, according to the matrix (\ref{O_matrix}).

    \begin{figure}
    \centering
    \includegraphics[width=1.05\linewidth]{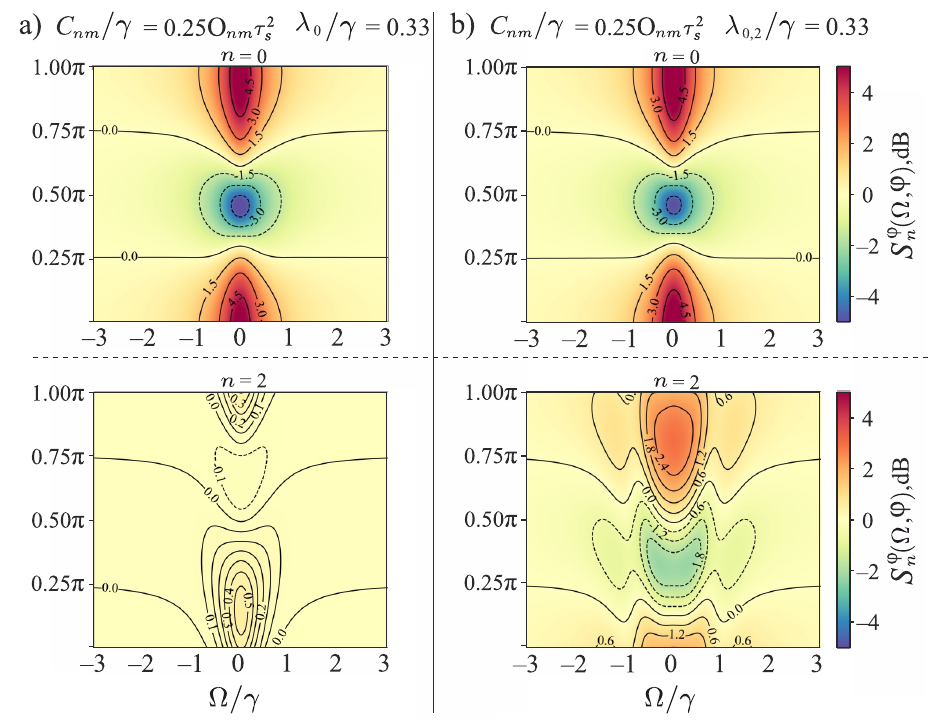}
    \caption{
    Modelled dependence of the photocurrent spectrum on the phase of the local oscillator in the scheme of balanced homodyne detection of a pulsed signal light generated in SPOPO for different operation regimes and different values of the second-order dispersion. (Left) The $0$th supermode is excited in SPOPO and the $0$th and the $2$nd supermodes are detected. (Right) The $0$th and the $2$nd supermodes are equally excited and detected.
    The chosen parameters correspond to the cases in Fig.~\ref{S_W}(c, g).}
    \label{fig_local_oscillator_phase}
    \end{figure}

Fig.~\ref{fig_local_oscillator_phase} shows dependence of the photocurrent spectrum on the phase of the local oscillator when the $0$th or $0$th and $2$nd modes are excited and detected.
Maximum shot noise suppression/excess is observed when the local oscillator phase takes values different from 0 and $\pi/2$. The actual phases depend on the frequency.
The effect is caused by the mode dependent detuning $\D_n$ in the presence of the dispersion (\ref{D_n_def}), which is more pronounced for higher order modes according to the diagonal elements of the overlap matrix for the Hermite-Gaussian modes (\ref{O_matrix}).

    \begin{figure}[ht]
    \centering
    \includegraphics[width=1.05\linewidth]{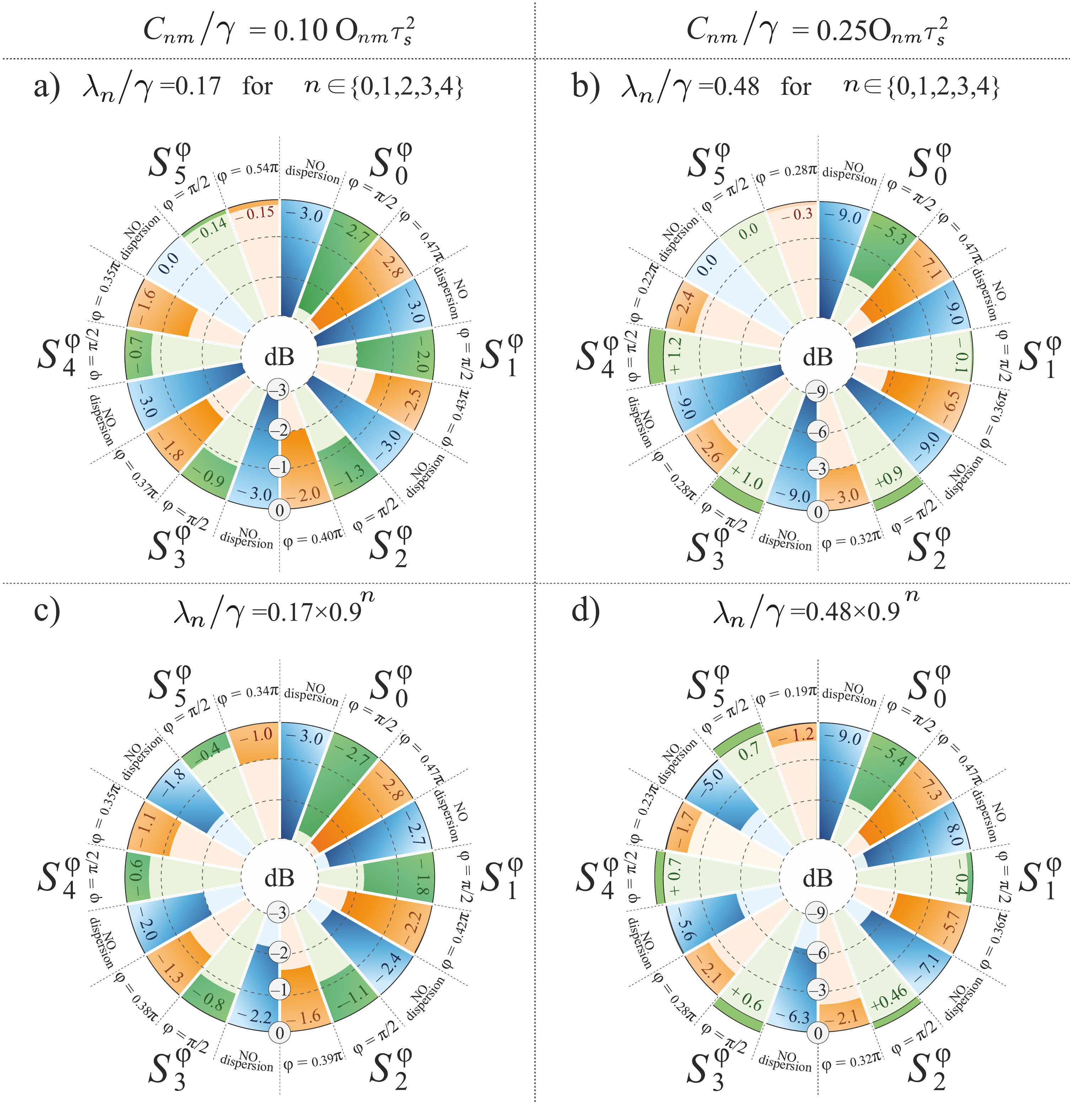}
    \caption{
    Simulated shot noise suppression of the difference photocurrent measured for the first five supermodes at zero side-band frequency in two regimes of SPOPO operation: (Top) Supermodes are equally excited in the SPOPO; (Bottom) Excitation of higher order supermodes decreases exponentially. The blue bars show noise suppression without the dispersion. The green and orange bars represent shot noise suppression in the presence of dispersion for the local oscillator phase equal to $\pi/2$ or the optimal value, respectively. The optimal phase of local oscillator is found in each case and depicted in the figure.
    } 
    \label{barplot}
    \end{figure}

Fig.~\ref{barplot} compares the suppression of shot noise for the first five supermodes at zero frequency without and with dispersion inside the cavity for two regimes of SPOPO operation: equal excitation of the first five supermodes, and excitation of the first five supermodes following a geometric progression dependence.
The results show that the maximal shot-noise suppression is reduced when dispersion is present, and it is observed at local oscillator phases different from $\pi/2$. The corresponding optimal values of the local oscillator phase are depicted in the figure.
The reduced noise suppression can be attributed to the entanglement of modes due to their coupling and the resulting a non-pure mixed state, as well as the mode-frequency dependent rotation of the squeezing ellipse of each mode's squeezed state.
This effect is more pronounced for stronger dispersion and higher excitation of the modes.
In particular, the difference $d_i$
 between the maximum shot-noise suppression without and with dispersion grows with the mode number, i.e. for $\lambda_n/\gamma=0.48$ (see Fig.~\ref{barplot}a) $d_i \in \{1.9,2.5, 6.0, 6.4, 6.6\}$dB and for $\lambda_n/\gamma=0.17$ (see Fig.~\ref{barplot}b) $d_i \in \{0.2,0,5, 1.0, 1.2, 1.4\}$,  where $i \in \{1,2,...,5\}$ dB.  Thus, the suppression of shot noise for the fundamental Gaussian supermode is more resistant to the influence of dispersion than any of the other Hermite-Gaussian supermodes.

 We note that in our SPOPO and homodyne detection model, the maximum shot noise suppression for all supermodes without dispersion is observed when the local oscillator has the same phase. This is due to the complex multiplier $i^n$ present in the definition of the supermode profiles (\ref{HG_def}), which leads to a phase difference of $\pi/2$ between the even and odd Hermite-Gaussian supermodes. 
 On the other hand, this can be understood as meaning that the phase of the local oscillator pulses must change by $\pi/2$ when detecting even/ odd modes in order for maximum shot noise suppression to be observed, which is consistent with the experiment \cite{Roslund2014}.

\section{Conclusion}
\L{Sec7}

We analyzed the effect of the non-compensated intracavity  group-velocity dispersion on the generation of temporally multimode squeezed pulsed light in SPOPO below the oscillation threshold.
We derived the Heisenberg-Langevin equation for the amplitude of the signal pulse inside the cavity taking into account several effects, in particular, multimode nature
of pulsed parametric down-conversion.
We show thath the dispersion causes a different phase to be accumulated along the cavity round trip for the broadband supermodes of the parametric process.
As a result, the cavity exhibits different resonance frequencies for each supermode, eliminating the frequency degeneracy that would otherwise be present in a dispersion-compensated cavity.
Stronger dispersion in a high-quality cavity leads to greater differences in resonance frequencies among the modes.

Decomposing the amplitude of the signal pulse in supermodes basis of the parametric process 
the system is equivalent to multiple parametric oscillators.
In this representation the dispersion causes the  mode-dependent detuning of each supermode oscillator from the cavity resonance and linear coupling between the oscillators.
Stronger dispersion and a higher-quality cavity lead to stronger coupling between the modes.
As a result, dispersion affects the quantum state of supermodes by impacting their squeezing level and the rotation of the axis of the squeezing ellipse. The mode coupling leads to entanglement among the original supermodes, resulting in the reduced squeezing level and state purity.
The system's operation regime is determined by the interplay between the rate of parametric photon pair generation in each supermode, the loss of photons through the cavity's coupling mirror, and the leakage of photons into coupled modes.
We considered an experimentally relevant operation regime, where the loss rate of photons in each supermode from the cavity is higher than their leakage to neighboring modes.
Using the perturbation theory up to the second order in coupling, we obtained the solution for amplitudes of supermodes for arbitrary number of modes, and pump level.
We further obtained the expression for the photocurrent spectrum measured in the time-domain balanced detection of supermodes
and analyzed suppression/enhancement (squeezing/anti-squeezing) of the shot noise of the photocurrent  for a set of parameters that correspond to several experimentally relevant cases.
Non-compensated intracavity dispersion leads to several effects: reduced detected squeezing/anti-squeezing compared to the dispersion-compensated case; deviation of the frequency dependence of squeezing from a Lorentzian function; maximum squeezing/anti-squeezing observed at a phase of the local oscillator different from $0$ and $\pi/2$, with the phase depending on frequency.
These effects are particularly noticeable for frequencies near zero and for higher order supermodes.
Proposed model of the system and the obtained solutions also suggests finding new supermodes in the presence of dispersion in pure independent squeezed states.
A recently developed theoretical approach of the "morphing supermodes" \cite{gouzien_morphing_2020} can be used.

The obtained in the work solutions for multiple linearly coupled parametric oscillators complement previous results obtained for a pair of coupled parametric oscillators \cite{olsen_entanglement_2005, roy_nondissipative_2021, jabri_light_2024}.
The study also reveals an analogy between SPOPO with group-velocity dispersion and coupled nonlinear waveguides \cite{barral_continuous-variable_2017, barral_quantum_2020}: the temporal evolution of coupled supermodes of the parametric process in SPOPO is similar to the spatial evolution of waveguide modes experiencing parametric amplification in each waveguide, distributed linear losses, and evanescent coupling between waveguides.
SPOPO with group-velocity dispersion can be used as a testbench for experimental study of multimode effects in coupled parametric oscillators.

\section{ACKNOWLEDGMENTS}
We thank Nicolas Treps and Claude Fabre for bringing the problem of SPOPO with the non-compensated intracavity group-velocity dispersion to our attention and for supporting the initial steps of the work. We also appreciate the discussions with Giuseppe Patera.
K.T. and D.M. acknowledge the financial support from the Russian
Science Foundation (Project No. 24-22-00318).

\appendix

\section{Derivation of the Heisenberg-Langevin equation for SPOPO}\label{AppendixA}

Here we derive a Heisenberg-Langevin equation for the amplitude of the optical signal pulse generated inside SPOPO upon its pump with the coherent pulsed pump light.
The derivation is similar to \cite{averchenko_quantum_2011, jiang_timefrequency_2012, jankowski_ultrafast_2024}.

An operator of the electric field of the signal light can be decomposed as follows: 
	\begin{align}
	E(t,z) = \frac{1}{\sqrt{2\pi}}\int a(\w,z) e^{-\ii\w t} \ud\w + h.c.
	\end{align}
Here $a(\w,z)$ is the operator of amplitude of a monochromatic signal mode at a given longitudinal coordinate $z$ along the propagation direction. We also consider a single spatial mode and neglect the dependence on transverse coordinates.
We consider a ring optical cavity of the length $L$. We assume one mirror of the cavity at $z=0$ is semitransparent with the reflection and transmission coefficients $\cal{R}$ and $\cal{T}$, respectively. The beamsplitter relation for amplitudes of signal fields on the mirror reads:
	\begin{align}
	a(\w,0) = \sqrt{\cal R} \; a(\w,L) + \sqrt{\cal T} \; a^\inn(\w,0)
    \L{BS_eq}
	\end{align}
We assume that the cavity is filled with transparent material with the second-order optical nonlinearity where SPDC of the pump pulsed light into signal pulses takes place. 
Then one can write the following general Bogolubov transformation that couples amplitudes of the signal field before and after the round trip inside the cavity filled with the nonlinear material \cite{wasilewski_pulsed_2006}:
	\begin{align}
	a(\w,L) = \int\ud\w' \[C(\w,\w')a(\w',0) + S(\w,\w')a^\+(\w',0)\]
	\end{align}
Coefficients of the transformation $C(\w, \w'), S(\w, \w')$ depend on the pump field, nonlinearity, geometry and dispersion of nonlinear material. 
We consider the weak nonlinear conversion limit, such that the probability of the generation of more than two signal photon pairs per round trip is negligible.
Then the transformation takes the form:
	\begin{align}
	a(\w,L) =  e^{\ii k(\w) L} \(  a(\w,0)  + \int \tilde S(\w,\w') a^\+(\w',0) \ud\w' \) 
    \L{Bogolubov_approx}
	\end{align}
Here the complex integral kernel $\tilde S(\w,\w')$ is symmetric and reads
    \begin{align}
    & \tilde S(\w,\w') \propto E_p(\w+\w') e^{\ii \frac{\Delta k(\w,\w') L}{2}} \text{sinc}\frac{\Delta k(\w, \w')L}{2},\\
    & \Delta k(\w, \w') = k_p(\w+\w') - k(\w) - k(\w'),
    \end{align}
where $E_p(\w)$ is the spectral amplitude of the pump pulse. Second line represents the phase mismatch between the pump and the pair of down-converted photons.

We define a field envelope with respect to half of the pump frequency $\w_0$	
	\begin{align}
	A(t,z) = \frac{1}{\sqrt{2\pi}}\int a(\w,z)e^{-\ii(\w-\w_0)t}\ud\w
	\end{align}
We consider  dispersion of the nonlinear material up to the second order as follows:
	\begin{align}
	k(\w) = k(\w_0) + k'(\w_0) (\w-\w_0) + \frac{k''(\w_0)}{2}(\w-\w_0)^2
	\end{align}
Then under an additional assumption of the small second-order dispersion, such that the series expansion is valid $e^{\ii \(1 + \ii \frac{k''(\w_0)L}{2} (\w-\w_0)^2\)} \approx 1 + \ii \frac{k''(\w_0)L}{2} (\w-\w_0)^2$ the transformation (\ref{Bogolubov_approx}) converts to the following transformation for the signal field envelope:
	\begin{multline}
	A(t,L) =  e^{\ii k(\w_0)L} \(1 - \ii \frac{k''(\w_0)L}{2} \frac{\partial^2}{\partial t^2}\) \\
    \times \(A(t-T_R,0) + \int K(t-T_R,t') A^\+(t',0) \ud t' \)
    \L{round_eq}
	\end{multline}
where: 
	\begin{align}
	& T_R = k'(\w_0) L,\\
    & K(t,t') =  \frac{1}{2\pi} \iint \tilde S(\w, \w') e^{-\ii (\w+\w') t} \ud \w \ud \w'
	\end{align}
We assume that the cavity can be off-resonant for the half pump frequency and make a series expansion:
	\begin{align}
    e^{\ii k(\w_0) L} = e^{\ii k(\w_0)(L_0 + \dd L)} \approx 1 + \ii k(\w_0) \dd L.
	\end{align}
Combining (\ref{BS_eq}) and (\ref{round_eq}) one gets for the signal field envelope on the coupling mirror
	\begin{multline}
	A(t,0) \approx (1 + \ii k(\w_0) \dd L) \sqrt{\cal R} A(t-T_R,0) \\
    - \ii \sqrt{\cal R} \frac{k''(\w_0)L}{2} \frac{\partial^2}{\partial t^2} A(t-T_R,0)\\
	+ \sqrt{\cal R}\int K(t-T_R,t') A^\+(t',0) \ud t' + \sqrt{\cal T} A^\inn(t,0)
	\end{multline}
We consider a difference of the field envelope after and before round-trip inside the cavity at time instants separated by the round-trip time $T_R$ divided by this time:
	\begin{multline}
	\frac{A(t,0)-A(t-T_R,0)}{T_R} \\
    = \(-\frac{1-\sqrt{\cal R}}{T_R} + \ii\sqrt{\cal R}\frac{k(\w_0) \dd L}{T_R}- \ii \sqrt{\cal R}\frac{k''(\w_0)L}{2 T_R} \frac{\partial^2}{\partial t^2}\)\\
    \times A(t-T_R,0) + \frac{\sqrt{\cal R}}{T_R}\int K(t-T_R,t') A^\+(t',0) \ud t' \\
    + \frac{\sqrt{T}}{T_R} A^\inn(t,0)
    \L{eq_difference}
	\end{multline}
Assuming that the field change after the round-trip is small we consider the above difference as an approximation to the first order derivative of a field envelope, which depends on additional time variable $T$:
	\begin{align}
	\frac{\partial}{\partial T}A(t,T)&=\frac{A(t,0)-A(t-T_R,0)}{T_R}
	\end{align}
Dependence on the variable $T$ describes the change of the envelope over time scales larger than the cavity round-trip time, similar to \cite{averchenko_quantum_2011, rana_quantum_2004, jankowski_ultrafast_2024} and $A(t,T=N T_R)$ gives the envelope after $N$ round trips.
We also define the quantities:
	\begin{align}
	\frac{\g}{2} &= \frac{1-\sqrt{\cal R}}{T_R},\\
	\Delta &= \sqrt{\cal R}\frac{k(\w_0) \dd L}{T_R},\\
	D &= \sqrt{\cal R}\frac{k''(\w_0)L}{2 T_R}
    \L{D_def}
    ,\\
	\frac{1}{2}G(t,t') &= \frac{\sqrt{\cal R}}{T_R} K(t,t'),\\
	\frac{\sqrt{\cal T}}{T_R} &\approx \sqrt{\frac{\g}{T_R}} ,\\
	F^\inn(t,T) &= \sqrt{\frac{\g}{T_R}} A^\inn(t,T)
	\end{align}
Then equation (\ref{eq_difference}) takes the form:
	\begin{multline}
	\frac{\partial}{\partial T}A(t,T) = \(-\frac{\g}{2} + \ii \Delta - \ii D \frac{\partial^2}{\partial t^2}\)A(t,T)\\
    + \frac{1}{2}\int G(t,t') A^\+(t',T) \ud t' + F^\inn(t,T)
	\end{multline}
The above derivation steps can be generalized to the case when the nonlinear medium occupies only a part of the space inside the cavity, as well as, there are additional transparent materials with the dispersion inside the cavity.
It results in the same equation with the modified parameters $T_R, \D, D$, which are the sums of the parameters for individual elements.

\section{Perturbation theory solution of intracavity coupled equations for supermodes.} \label{AppendixB}

We solve coupled equations (\ref{coupled_modes_equations}) using the perturbation theory with respect to the coupling coefficients of modes $C_{nm}$ as follows.

\subsubsection{$0$th order solution}

Setting $C_{nm} = 0$ one gets a set of independent equations in the zeroth order of the perturbation theory:
	\begin{multline}
	-\ii\W a_n^{(0)}(\W) \\
    = \(-\frac{\g_n}{2} + \ii\Delta_n\)  a_n^{(0)}(\W) 
    + \frac{\l_n}{2} a_n^{(0)\+}(-\W) +  f_n^{(0)}(\W),
	\L{0th_coupled_modes_equations}
	\end{multline}
where
	\begin{align}
    f_n^{(0)}(\W) = f_n(\W)
	\end{align}
the zeroth order solution can be represented as follows
	\begin{align}
	& a_n^{(0)}(\W) = U_n(\W) f_n^{(0)}(\W) + V_n(\W) f_n^{(0)\+}(-\W),
    \L{0th_sol}
	\end{align}
where
	\begin{align}
	& U_n(\W) = \frac{1}{H_n(\W)},\\
	& V_n(\W) = \frac{1}{H_n(\W)} \frac{\l_n}{2} \frac{1}{\frac{\g_n}{2}+\ii \D_n - \ii \W},\\
	& H_n(\W) = \frac{\g_n}{2}-\ii \D_n - \ii \W - \(\frac{\l_n}{2}\)^2 \frac{1}{\frac{\g_n}{2}+\ii\D_n-\ii\W}
    \L{}	
	\end{align}

\subsubsection{1-st order solution}

Second, we substitute the solution (\ref{0th_sol}) in equation (\ref{0th_coupled_modes_equations}) for the term describing the linear coupling of modes.
One gets the equation for the first order which is identical to the equation for the zero order (\ref{0th_coupled_modes_equations}) with the modified noise term:
	\begin{multline}
	-\ii\W a_n^{(1)}(\W) \\
    = \(-\frac{\g_n}{2} + \ii\Delta_n\) a_n^{(1)}(\W) + \frac{\l_n}{2} a_n^{(1)\+}(-\W) + f_n^{(1)}(\W),
	\L{1st_coupled_modes_equations}
	\end{multline}
where the modified noise term reads 
	\begin{align}
	f_n^{(1)}(\W) = f_n^{(0)}(\W) - \ii\sum_{m\neq n} C_{nm} a_m^{(0)}(\W).
	\end{align}
The first-order solution has the same form as the zero-order solution (\ref{0th_sol})
	\begin{align}
	& a_n^{(1)}(\W) = U_n(\W) f_n^{(1)}(\W) + V_n(\W) f_n^{(1)\+}(-\W),
	\end{align}
Or in terms of the original noise $f_n^{(0)}(\Omega)$:
\begin{equation}
    a_n^{(1)}(\Omega) = a_n^{(0)}(\Omega) + \delta a_n^{(1)}(\Omega),
    \label{a1}
    \end{equation}
where
\begin{equation}
    \delta a_n^{(1)}(\Omega) = \sum_{m \neq n} \left[U_{nm}(\Omega) f_m^{(0)}(\Omega) + V_{nm}(\Omega) f_m^{(0)\dagger}(-\Omega)\right],
\end{equation}
and
\begin{equation}
    \begin{aligned}
        U_{nm} (\Omega) &= i \left[
        - U_n(\Omega) C_{nm} U_m(\Omega) + V_n(\Omega) C_{nm}^* V_m^*(-\Omega)\right],\\
        V_{nm} (\Omega) &= i \left[ - U_n(\Omega) C_{nm} V_m(\Omega) + V_n(\Omega) C_{nm}^* U_m^*(-\Omega)\right].
    \end{aligned}
\label{uv}
\end{equation}

\subsubsection{$2$nd order solution}

We repeat the above procedure to get the second-order perturbation solution that formally reads
	\begin{align}
	& a_n^{(2)}(\W) = U_n(\W) f_n^{(2)}(\W) + V_n(\W) f_n^{(2)\+}(-\W),\\
	& f_n^{(2)}(\W) = f_n^{(0)}(\W) - \ii\sum_{m\neq n} C_{nm} a_m^{(1)}(\W). 
	\end{align}
In terms of $f_n^{(1)}(\W)$, analogically to (\ref{a1}):
\begin{equation}
    a_n^{(2)}(\Omega) =  a_n^{(0)}(\Omega) + \Delta a_n^{(2)}(\Omega),
    \end{equation}
where
\begin{align}
    & \Delta a_n^{(2)}(\Omega) = \sum_{m \neq n} \left[U_{nm}(\Omega) \hat f_m^{(1)} (\Omega) + V_{nm}(\Omega) \hat f_m^{(1)\dagger} (-\Omega)\right], \\
    & f_m^{(1)}(\W) = f_m^{(0)}(\W) - \ii\sum_{p\neq m} C_{mp} a_p^{(0)}(\W).
\end{align}
In terms of the original noise $f_n^{(0)}(\Omega)$:
\begin{align}
    a_n^{(2)}(\Omega) = a_n^{(0)}(\Omega) + \delta a_n^{(1)}(\Omega) + \delta a_n^{(2)}(\Omega),
    \label{a2}
    \end{align}
where
\begin{equation}
    \delta a_n^{(2)}(\Omega) = \sum_{\substack{m \neq n \\ p \neq m}} \left[U_{nmp}(\Omega) f_p^{(0)}(\Omega) + V_{nmp}(\Omega) f_p^{(0)\dagger}(-\Omega)\right],
\end{equation}
and
\begin{equation}
    \begin{aligned}
        {U_{nmp}(\Omega)} &= i [-U_n(\Omega) C_{nm} U_{mp}(\Omega) + V_n(\Omega) C_{nm}^* V_{mp}^*(-\Omega)],
        \\
        {V_{nmp}(\Omega)}  &= i [-U_n(\Omega) C_{nm} V_{mp}(\Omega) + V_n(\Omega) C_{nm}^* U_{mp}^*(-\Omega)].
    \end{aligned}
\end{equation}

\section{Calculation of the photocurrent spectrum of balanced homodyne detection}\label{AppendixC}

The photocurrent spectrum reads according to the definition (\ref{II2_total}) and expression (\ref{Iw_BHD_def}):
\begin{multline}
    S_{n}^\varphi(\W) = \<I_n^\varphi(\W) I_n^{\varphi\+}(\W)\>\\
    \propto \<b_n(\W) b_n^\+(\W)\> + \<b_n^\+(-\W) b_n(-\W)\> 
    \\ + \<b_n(\W) b_n(-\W)\> e^{\ii2\varphi} + \<b_n^\+(-\W) b_n^\+(\W)\> e^{-\ii2\varphi} 
    \L{S_derivation}
\end{multline}

Further we calculate corresponding correlators of the output field amplitude using the solution for the output field on the $n$-th mode (\ref{b_solution}) and the only non-zero correlation function for the vacuum noise (\ref{f_correlator}).
We calculate the correlators up to the second-order with respect to the coupling coefficient between modes $C_{nm}$.
The correlators read:
\begin{multline}
    \langle b_n(\Omega) b_n^\dagger (\Omega') \rangle \\
    \propto \g_n |W_n(\W)|^2 + \sum_{m \neq n} \g_m |U_{nm}(\W)|^2 
    \\
    + \g_n\sum_{m\neq n} \(W_n(\W) U_{nmn}^*(\W) + \text{c.c.}\)
\end{multline}
\begin{multline}
    \< b_n^\+(-\W) b_n (-\W') \> \\
    \propto \g_n |V_n(\W)|^2 + \sum_{m \neq n} \g_m |V_{nm}(\W)|^2 
    \\
    + \g_n\sum_{m\neq n} \(V_n(-\W) V_{nmn}^*(-\W) + \text{c.c.}\)
    \L{b2}
\end{multline}
\begin{multline}
    \< b_n(\W) b_n (-\W') \> \\
    \propto \g_n W_n(\W) V_n(-\W) + \sum_{m \neq n} \g_m U_{nm}(\W) V_{nm}(-\W) 
    \\
    + \g_n\sum_{m\neq n} \(W_n(\W) V_{nmn}(-\W) + V_n(-\W) U_{nmn}(\W)\)
\end{multline}
\begin{align}
    \< b_n^\+(-\W) b_n^\+ (\W') \>  = \< b_n(\W) b_n (-\W') \>^*
\end{align}
The proportionality symbol is used because we are omitting the term $\g_n T_R \dd(\W-\W')$ in all expressions.
Substituting these correlotors in the expression (\ref{S_derivation}) and performing simplifications one gets the expression (\ref{II2_total}).

    \begin{figure}[t]
    \centering
    \includegraphics[width=\columnwidth]{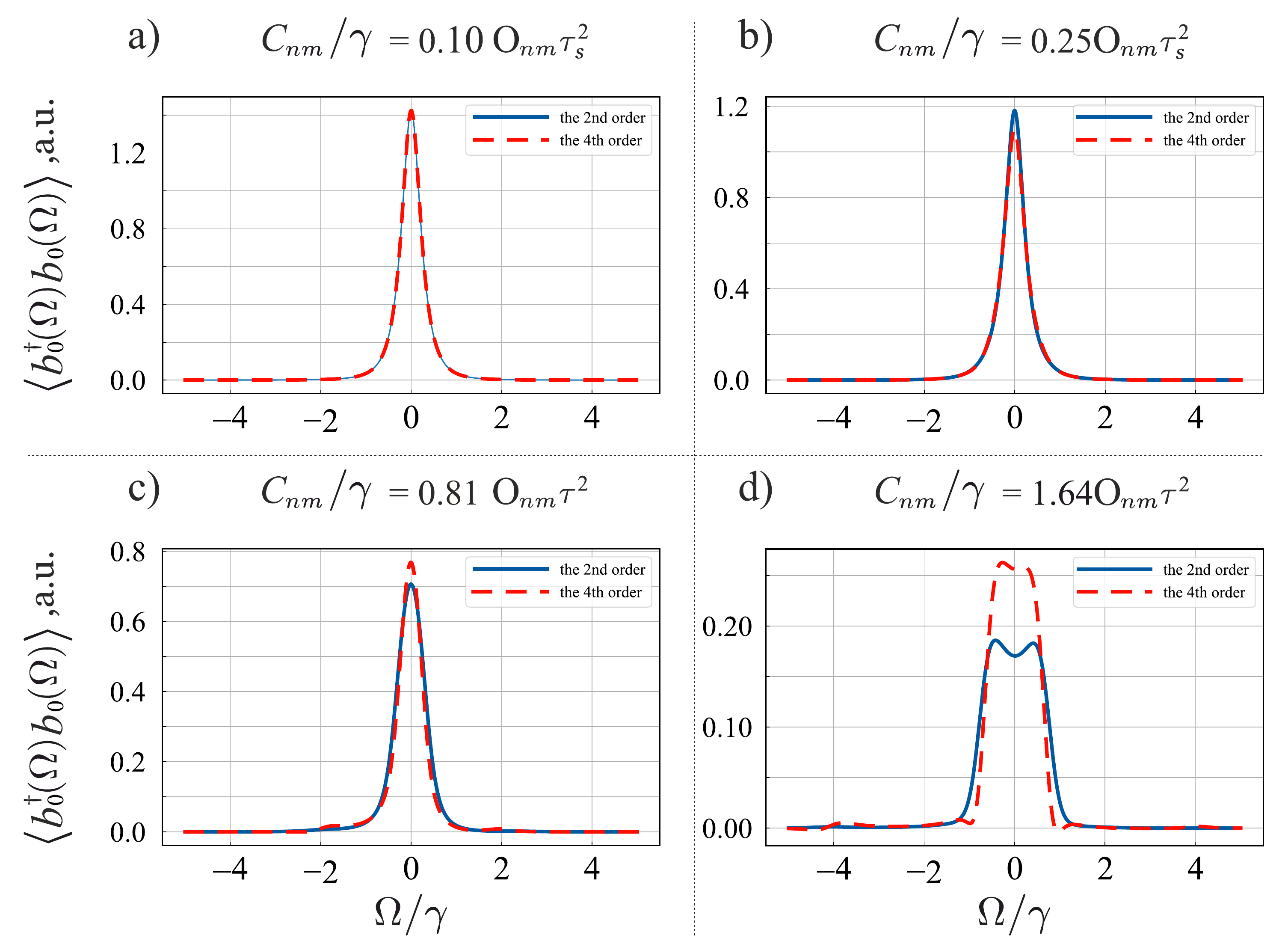}
    \caption{The spectrum of the amplitude of the $0$th supermode oscillator $\langle b_0^\dagger(\Omega) b_0 (\Omega) \rangle $ obtained with the perturbation theory up to the $2$nd order (the solid blue line) and up to the $4$th order (the dashed red line) at the dimensionless pump rate $\lambda_0/\gamma = 0.48$.
    } 
    \label{fig2vs4}
    \end{figure}

    \section{
The spectrum of the amplitude of the $0$th supermode oscillator calculated with the perturbation theory up to the $4$th order in the coupling coefficients }\label{AppendixD}

To verify the numerical results in Sec.~\ref{Sec6} and validity of the approximations used, one needs to estimate some physical observable obtained with the perturbation theory up to the next nonzero order. In particular, this will allow one to see whether the $2$nd order of the perturbation theory is enough for the given values of the coupling coefficients between modes $C_{nm}$ or  the next nonzero order of the perturbation theory must be taken into account. We note, that all expressions in any order of the perturbation theory can be derived in the same way as described in Sec.~\ref{Sec4}.

 Fig.~\ref{fig2vs4} shows the quantity $\langle b_0^\dagger(\Omega) b_0 (\Omega) \rangle$ (\ref{b2}) which represents the spectrum of the amplitude of the $0$th supermode oscillator obtained with the perturbation theory up to the $2$nd order and up to the $4$th order at the same dimensionless pump rate $\lambda_0/\gamma = 0.48$. We choose this observable due to its non-negativity for any values of its argument 
 and  ease of calculation.  As one can see, at low values of the dispersion strength (i.e. the coupling coefficients) (Fig.~\ref{fig2vs4}a and Fig.~\ref{fig2vs4}b)
the results almost perfectly coincides,  showing that the $2$nd order is enough to describe the behvaiour of the system. At the same time, high values of the dispersion strength (Fig.~\ref{fig2vs4}c and Fig.~\ref{fig2vs4}d) demonstrates some differences in the shape of the curves. Moreover, in the Fig.~\ref{fig2vs4}d one can also see the negative non-physical values of the chosen observable. This signifies that the next nonzero order of the perturbation theory should be taken into account.  

\end{document}